\documentclass[3p,twocolumn]{elsarticle}

\usepackage{lineno,hyperref,pifont,natbib,geometry,fleqn,graphicx,makeidx,fancyhdr,multirow,verbatim,amsmath,amsthm,amssymb,float,array,subfloat,epstopdf}
\modulolinenumbers[5]

\journal{Journal of \LaTeX\ Templates}

\begin{document}

\begin{frontmatter}

\title{Processing and characterization of epitaxial GaAs radiation detectors} 

\author[a]{X. Wu\corref{cor1}}
\ead{Xiaopeng.Wu@vtt.fi}
\author[b]{T. Peltola\corref{cor2}}
\ead{timo.peltola@helsinki.fi}
\author[b]{T. Arsenovich}
\author[a]{A. G\"{a}dda} 
\author[b]{J. H\"{a}rk\"{o}nen} 
\author[c]{A. Junkes}
\author[b]{A. Karadzhinova} 
\author[d]{P. Kostamo}
\author[e]{H. Lipsanen} 
\author[b]{P. Luukka} 
\author[e]{M. Mattila} 
\author[d]{S. Nenonen} 
\author[a]{T. Riekkinen} 
\author[b]{E. Tuominen} 
\author[b]{A. Winkler}

\address[a]{VTT, Microsystems and Nanoelectronics, Tietotie 3, Espoo, P.O. Box 1000, FI-02044 VTT, Finland}
\address[b]{Helsinki Institute of Physics, P.O. Box 64 (Gustaf H\"{a}llstr\"{o}min katu 2) FI-00014 University of Helsinki, Finland}
\address[c]{Institute for Experimental Physics, University of Hamburg, Germany} 
\address[d]{Oxford Instruments Analytical Oy, Tarvonsalmenkatu 17, Espoo, Finland} 
\address[e]{Aalto University, Department of Micro and Nanosciences, Tietotie 3, Espoo, P.O. Box 13500, FI-02015, Finland}

\cortext[cor1]{Corresponding author}
\cortext[cor2]{Speaker, Corresponding author}

\begin{abstract}
\label{Abstract}
GaAs devices have relatively high atomic numbers (Z=31, 33) and thus extend the X-ray absorption edge beyond that of Si (Z=14) devices. In this study, radiation detectors were processed on GaAs substrates with 110 $\mu\textrm{m}$ - 130 $\mu\textrm{m}$ thick epitaxial absorption volume. Thick undoped and heavily  doped p$^+$ epitaxial layers were grown using a custom-made horizontal Chloride Vapor Phase Epitaxy (CVPE) reactor, the growth rate of which was about 10 $\mu\textrm{m}$/h. The GaAs p$^+$/i/n$^+$ detectors were characterized by Capacitance Voltage ($CV$), Current Voltage ($IV$), Transient Current Technique (TCT) and Deep Level Transient Spectroscopy (DLTS) measurements. The full depletion voltage ($V_{\textrm{\small fd}}$) of the detectors with 110 $\mu\textrm{m}$ epi-layer thickness is in the range of 8 V - 15 V and the leakage current density is about 10 nA/cm$^2$. 
The signal transit time determined by TCT is about 5 ns when the bias voltage is well above the value that produces the peak saturation drift velocity of electrons in GaAs at a given thickness. Numerical simulations with an appropriate defect model agree with the experimental results.

\end{abstract}


\end{frontmatter}


\section{Introduction}
\label{Introduction}
\begin{figure}[tbp] 
\centering
\includegraphics[width=.45\textwidth]{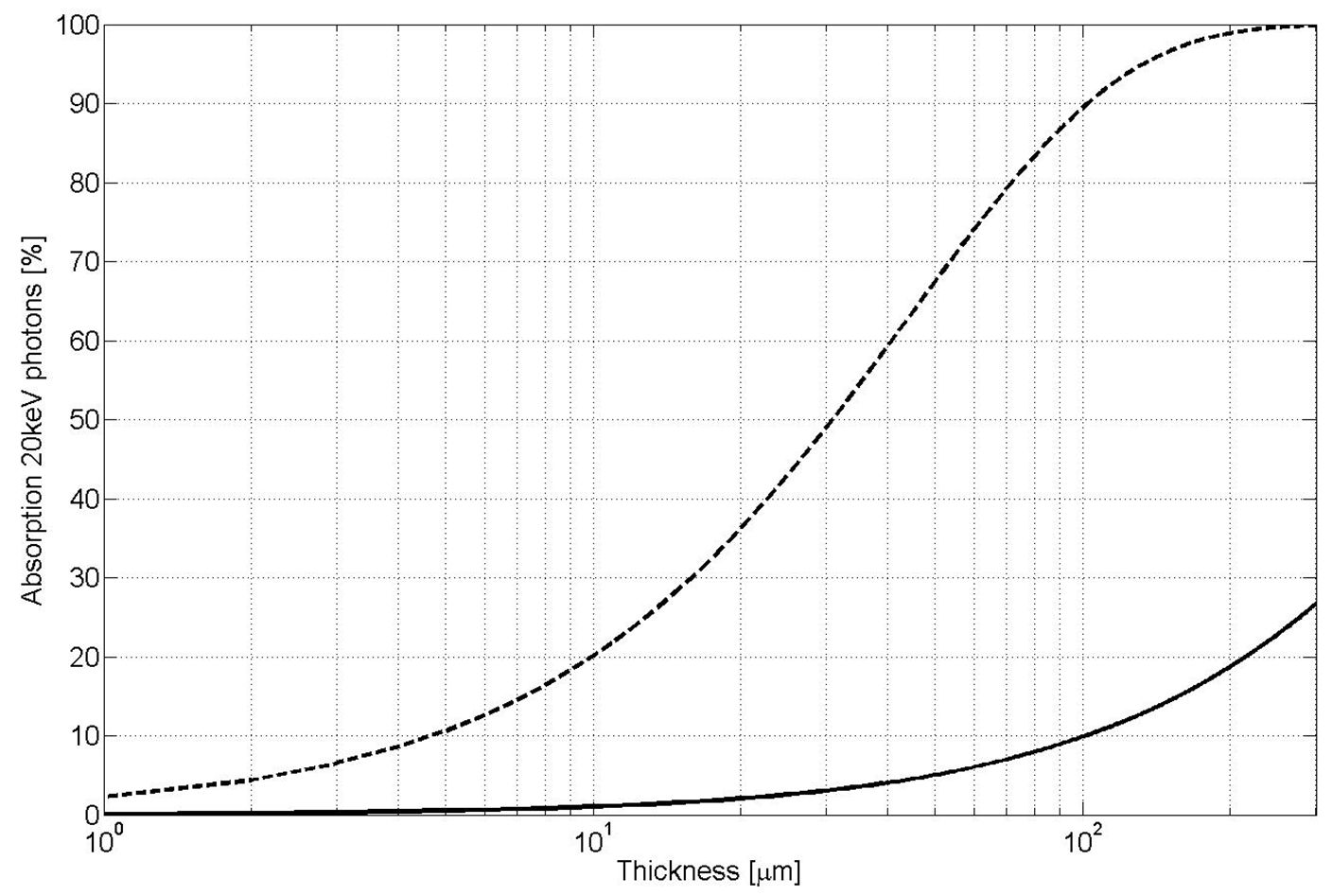}
\caption{Absorption efficiency vs. detector thickness calculated by Beer-Lambert law. Dashed and solid curves represent GaAs and silicon, respectively.} 
\label{fig:1a}
\end{figure}

Radiation detectors made on epitaxial GaAs are a promising alternative for the silicon devices used for spectroscopy and radiography applications requiring moderate photon energies, i.e. more than 10 keV. Mammography is an important example of such an application with large technological and societal impact. Earlier studies on using GaAs material for X-ray registration include References \cite{Owens2004,Sellin2006}. The atomic numbers of GaAs are 31 (Ga) and 33 (As) while the atomic number for Si is 14. As a result, for a low energy X-ray process with a photon energy of 20 keV, the mass attenuation coefficients ($\mu/\rho$) are 42.3 and 4.4 $\textrm{cm}^2$/g for GaAs and Si, respectively. The benefit of GaAs is well illustrated in figure~\ref{fig:1a}. Here, the total absorption efficiency of 20 keV photons, calculated by Beer-Lambert law, is shown as the function of detector thickness. 

As seen in figure~\ref{fig:1a}, the GaAs thickness of 100 $\mu\textrm{m}$ is enough to absorb about 90$\%$ of the 20 keV photons, while the absorption efficiency of 300 $\mu\textrm{m}$ thick Si would be only about 27$\%$. 

Being the basic starting material in optoelectronics industry, the processing technology of GaAs devices is well established. The epitaxy based on ultra-pure gaseous precursors is a pronounced approach to fabricate pin-diode structured GaAs radiation detectors. When illuminated by X-rays, a GaAs detector operates under the condition of moderate injection level, i.e. the carrier concentration is about $10^{17}$ $\textrm{cm}^{-3}$ \cite{Bhatta1997}. 
According to the well-known transferred electron effect \cite{Ruch1968,Dargys1994,Sze1981} (further described in section~\ref{TCAD Simulations}), the drift velocity of charge carriers in GaAs decreases after the electric field inside the material has reached the value of about 3.3 kV/cm \cite{Ruch1968}. In a 100 $\mu\textrm{m}$ thick GaAs detector, this saturation drift velocity of charge carriers would take place when the value of reverse bias is 33 V. The relatively low full depletion voltage ($V_{\textrm{fd}}$) of GaAs defines the maximum doping concentration to be about $5\times10^{11}$ $\textrm{cm}^{-3}$. This is challenging to achieve in GaAs crystals manufactured using the bulk growth methods.  

A number of deposition techniques have been developed for high purity GaAs epitaxy. Many of these epitaxial growth techniques, however, suffer from high cost, low growth rate or incapability to grow thick layers. On the other hand, chloride vapor phase epitaxy (CVPE) technique has been proved to grow high purity epi-GaAs with a high growth rate (about 10 $\mu\textrm{m}$/hour) and with an achievable layer thickness of more than 100 $\mu\textrm{m}$, as is required in detector applications. In the following, we present the processing, the results of characterization and the device simulations of the 110 $\mu\textrm{m}$ thick epitaxial pin-diode GaAs radiation detectors. 
%

\section{Processing and design of detectors}
\label{Processing of GaAs detectors}
Two 2-inch diameter GaAs wafers with CVPE grown intrinsic epitaxial layers were processed to radiation detectors at Micronova nanofabrication center cleanroom. The wafers used in the study were produced by A. F. Ioffe Physico-Technical Institute (Ioffe PTI) \cite{Kostamo2008}. The n-type 460 $\mu\textrm{m}$ thick “epi-ready” GaAs wafers were doped with silicon which had a concentration of about $10^{19}$ $\textrm{cm}^{-3}$. 
Undoped epitaxial layers, of thicknesses 110 $\mu\textrm{m}$ and 130 $\mu\textrm{m}$, were grown on the substrates in a horizontal CVPE reactor at 730-760 $^{\circ}$C temperatures. 
Epitaxy was finalized by the deposition of a 1-2 $\mu\textrm{m}$ thick zinc doped p$^+$ layer on the undoped layer to complete the desired p-i-n structure. 

\begin{figure}[tbp] 
\centering 
\includegraphics[width=.4\textwidth]{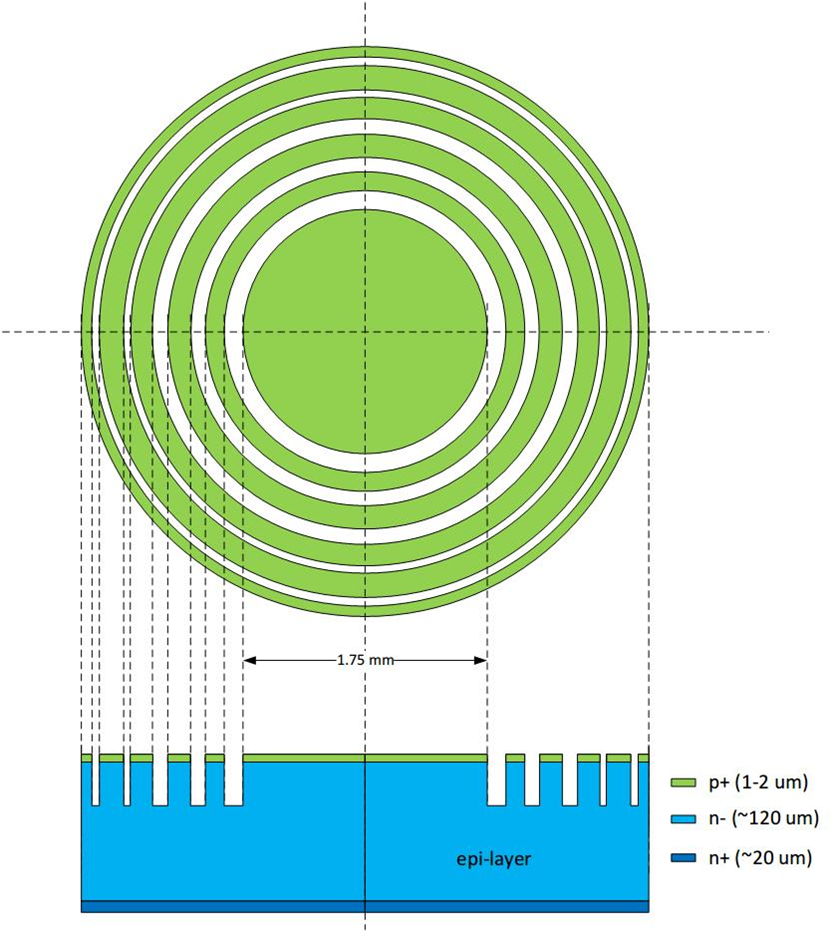}
\caption{Cross section and design of GaAs pin-diode.}
\label{fig:5}
\end{figure}

After the epitaxy, the wafers were cleaned with acetone and isopropanol to remove particles and possible organic contamination from the surfaces. 
Mesa structures shown in figure~\ref{fig:5} were formed by deep reactive ion etching (DRIE) with BCl3 plasma. The trench depth was roughly 7 $\mu\textrm{m}$ in order to ensure sufficient isolation through p$^+$ region between the segmented region. After etching, a plasma enhanced chemical vapor deposition (PECVD) grown dielectric layer was deposited for surface protection. Openings for contacts to PECVD dielectric were done by wet etching with diluted buffered hydrofluoric acid (BHF). The contact metallization was made by sputtered Ni/TiW/Ni (20 nm/20 nm/40 nm) stacked layers and patterning was completed by a lift-off process. The mechanical grinding and polishing were applied to the substrate side leaving a 20 $\mu\textrm{m}$ heavily doped n-type GaAs layer for the backside metal contact. As the last process step, a 100 nm thick Ni layer was sputtered on the back plane as the backside contact.

The wafer layout contains two pixel sensors compatible with CERN Medipix CMOS read-out ASIC \cite{Kostamo2008,Tlustos2008} ($256\times256$ pixels with 55 $\mu\textrm{m}$ pixel pitch), one strip sensor (128 strips with 200 $\mu\textrm{m}$ pitch) and several pad diodes (diameter of 1.75 mm). First results of the fabricated GaAs flip-chip bonded pixel devices and their response to X-ray illumination has been reported in Reference \cite{Wu2014}. In this report, we focus on diodes, of which design and dimensions are shown in figure~\ref{fig:5}. 

\section{Characterization}
\label{Detector Characterization}
\subsection{$CV$ and $IV$ results}
\label{$CV$ and $IV$ results}
\begin{figure}[tbp] 
\centering
\includegraphics[width=.45\textwidth]{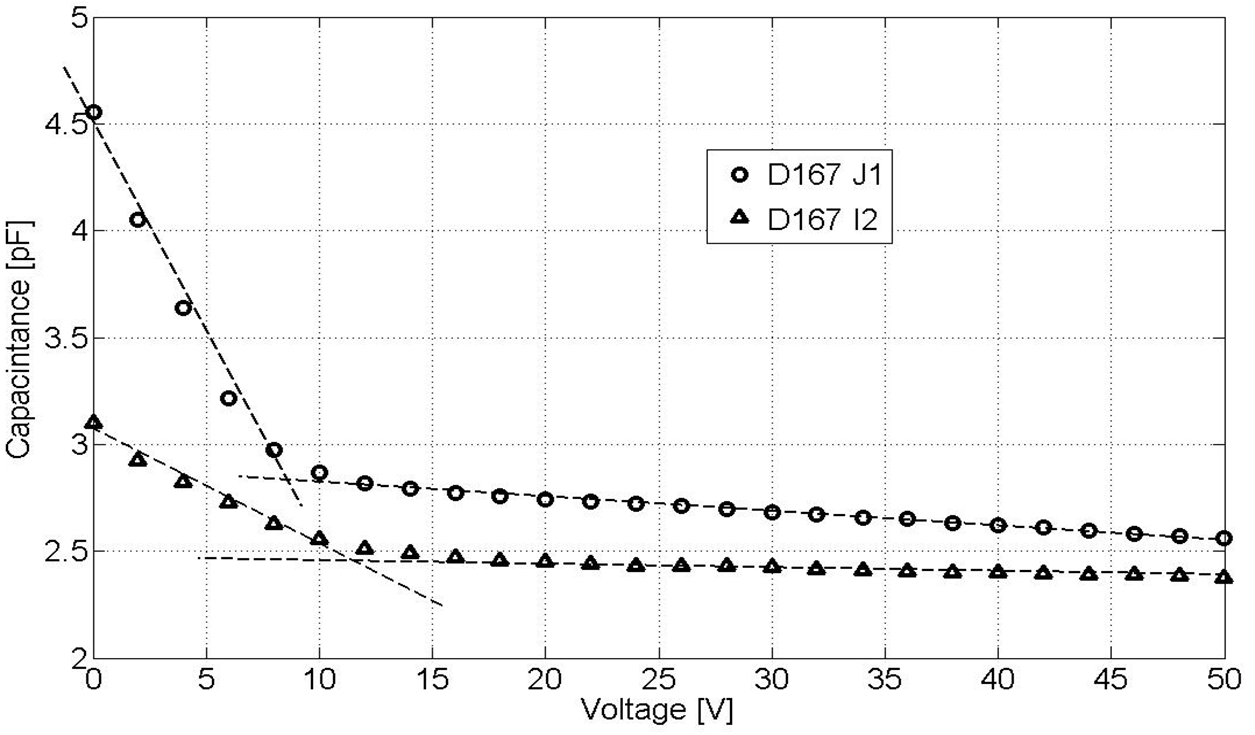}
\caption{$CV$ curves of the two (110 $\mu\textrm{m}$ thick i-layer) GaAs pad detectors. Geometrical capacitance calculated from the surface area in figure~\ref{fig:5} is $\sim$2.32 pF. The minimum capacitances indicate about 9$\%$ smaller active thickness for the D167 J1 diode. Full depletion voltages of $\sim$8.5 V and $\sim$12 V for the D167 J1 and D167 I2, respectively, were determined from the crossing points of the linear fits.} 
\label{fig:6a}
\end{figure}

\begin{figure}[tbp] 
\centering
\includegraphics[width=.45\textwidth]{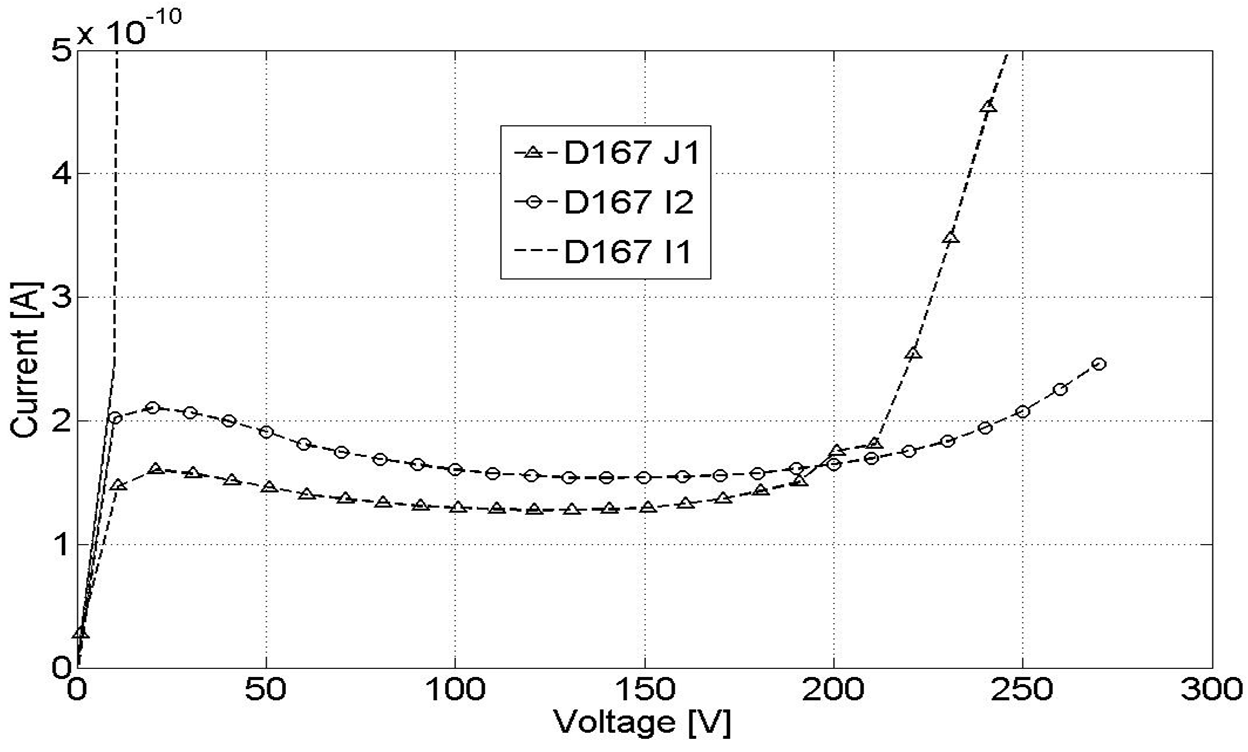}
\caption{Current Voltage curves of two (110 $\mu\textrm{m}$ thick i-layer) GaAs pad detectors.} 
\label{fig:6b}
\end{figure}

The pad detectors and test structures were characterized by Capacitance Voltage ($CV$) / Current Voltage ($IV$), Transient Current Technique (TCT) \cite{Eremin1996TCT} and Deep Level Transient Spectroscopy (DLTS) measurements. The full depletion voltage ($V_{\textrm{fd}}$) of the detectors is about 10 V and the leakage current at $V_{\textrm{fd}}$ was in most of the devices below 10 nA/$\textrm{cm}^2$. Some of the devices showed breakdown before the $V_{\textrm{fd}}$. Illustrative examples of $CV$ and $IV$ curves are shown in figures~\ref{fig:6a} and \ref{fig:6b}. 

The $CV$ measurements were done at room temperature at a frequency of 10 kHz. This was deemed sufficient since the low level of the leakage current density rules out the possibility of distortion from the generation current of the deep defect levels to the $CV$ results. It can be seen in figure~\ref{fig:6a} that the GaAs diodes do not show sharp capacitance saturation, which is often the case in e.g. planar silicon pad detectors. This is most likely due to non-abrupt backplane of the pin-diode. The epitaxial deposition at about 760 $^{\circ}$C of $\textgreater$100 $\mu\textrm{m}$ thickness takes more than ten hours. It is comprehensible that diffusion from heavily doped substrate takes place in some extent, resulting in a non-abrupt i-n$^+$  transition region that is observed by non-saturated $CV$ characteristics. The leakage currents in the devices were typically 100-200 pA, corresponding to the current density of about 10 nA/$\textrm{cm}^2$. The leakage current tends to saturate after the $V_{\textrm{fd}}$ and breakdown occurs in good devices after about 200 V, i.e. over ten times higher voltage than the $V_{\textrm{fd}}$. More statistics of the leakage current has previously been reported in Reference \cite{Wu2014}. 
%
\subsection{TCT results}
\label{Transient current technique results}
Transient current technique (TCT) measurements were performed at the premises of Detector Laboratory of Helsinki Institute of Physics (HIP). The detectors were assembled on a PCB substrate by a conductive silicon carbide tape and a probe needle. Optical excitation was done by a red laser (678 nm) emitting 30 ps (FWHM) pulse to the segmented p$^+$ side of the diode. Detectors were capacitively decoupled from the 4 GHz oscilloscope by Miteq Ltd. Bias-T and the signal was amplified by a broad band amplifier provided by Particulars Ltd. Since the absorption depth of 678 nm light in GaAs is less than 1 $\mu\textrm{m}$, the holes excited by the incident photons will be inherently quickly collected by the adjacent p$^+$-electrode and the current transient observed by the oscilloscope is the current created by the electrons drifting the entire thickness of the pin-diode. Since the epi-layer is n$^{--}$ type, the electrons are majority carriers (not minority) and their radiative lifetime can be ignored. In high purity, high quality GaAs the radiative lifetime can be very long, even in the microsecond regime \cite{Dingle1969}. 

\begin{figure}[tbp] 
\centering
\includegraphics[trim=0.05cm 0.05cm 0.05cm 0.05cm, clip=true, width=.48\textwidth]{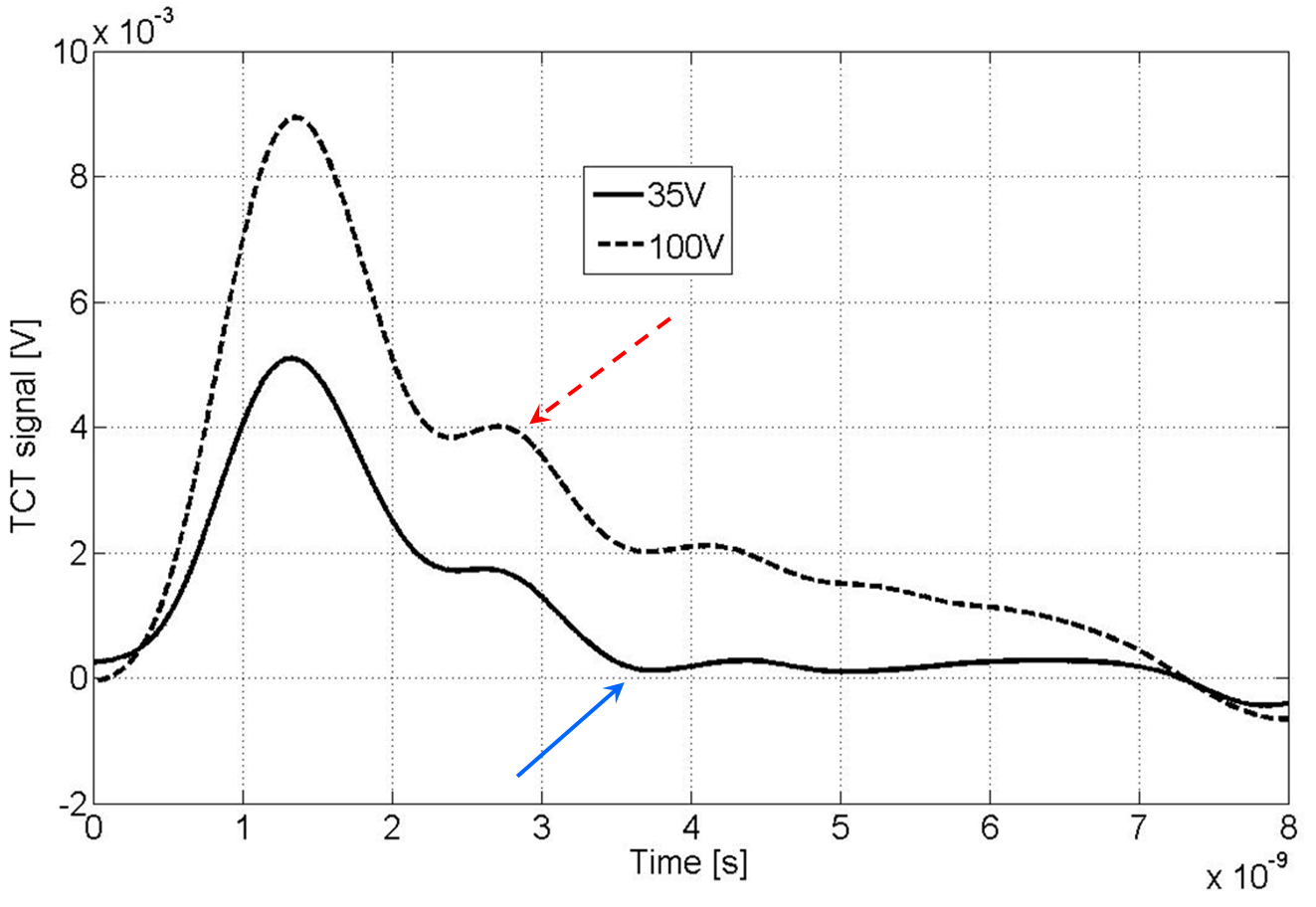}
\caption{TCT measurement for GaAs diode biased with 35 V and 100 V.} 
\label{fig:7}
\end{figure}

The curves shown in figure~\ref{fig:7} are the electron current transients at 35 V and 100 V bias voltages. The duration of the transient is about 5 ns when biased with 100 V. 
The electron current transient time at 35 V bias is clearly shorter, about 3.5 ns, than at 100 V indicating higher drift velocity in case of lower electric field. The parasitic oscillations after the second peak in both of the curves were not considered as a part of the signal. Thus, for the collection time estimation the slope of the 100 V curve was extrapolated to zero after the second peak (indicated by a red dashed arrow in the figure) and the duration of the 35 V curve was determined from the location indicated by a blue arrow. In this particular measurement, the laser spot was not collimated. Thus, there is discrepancy in integrated collected charge due to the charge flow in lateral direction. 

\subsection{DLTS results}
\label{Deep level transient spectroscopy results}
\begin{figure}[tbp] 
\centering
\includegraphics[width=.45\textwidth]{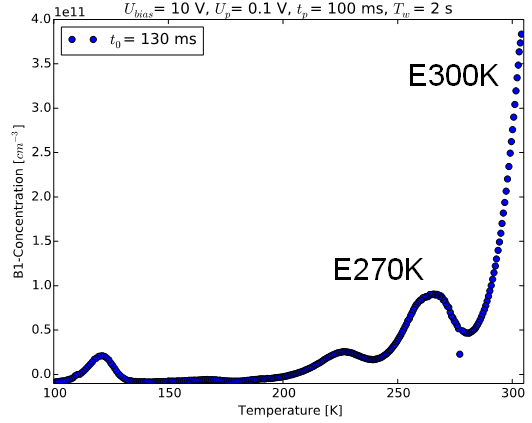}
\caption{DLTS measurement with electron injection. $U_{\textrm{bias}}$ is the reverse bias voltage, $U_{\textrm{p}}$ is the forward pulse voltage, $t_{\textrm{p}}$ is the pulse time, $T_{\textrm{w}}$ is the acquisition time window and $t_{\textrm{0}}$ data recording cut.} 
\label{fig:8}
\end{figure}

\begin{figure}[tbp] 
\centering
\includegraphics[width=.45\textwidth]{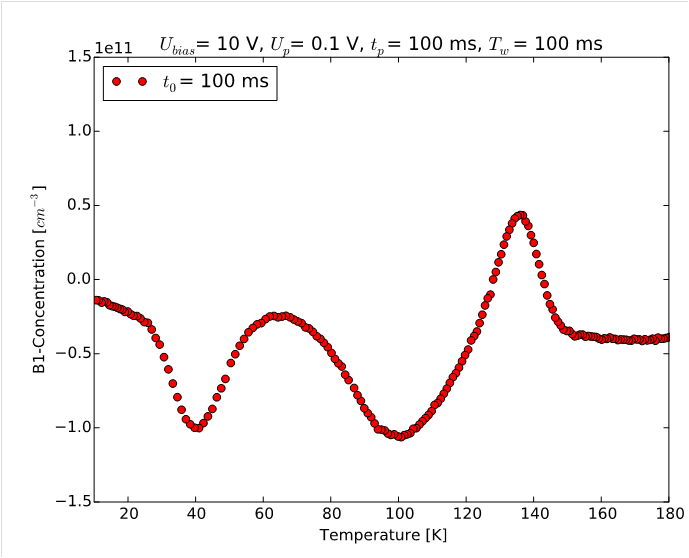}
\caption{DLTS measurement with hole injection.} 
\label{fig:8b}
\end{figure}

The Deep Level Transient Spectroscopy (DLTS) analysis for GaAs diodes was performed at University Hamburg. The DLTS spectra of electron and hole injections are shown in figures~\ref{fig:8} and \ref{fig:8b}, respectively. The analysis shows two deep electron traps peaking at $T$ = 270 K (E270) and $T$ $\geq$ 300 K (E300). Due to the technical restrictions, it was 
not possible to unambiguously define the electrical properties and concentration of E300 trap. 
The hole traps were observed at low temperatures, therefore they appear to be shallow defects, with concentrations of a few $10^{10}$ cm$^{-3}$. A contribution of such shallow defects to the leakage current is very unlikely. It is more apparent that deep defects contribute in GaAs pin-diodes. The deeper defects found in the DLTS spectra are better candidates for dominating the current. DLTS measurements up to 450 K have to be performed to check the existence of the very deep defects.
%

\section{TCAD simulations and modeling}
\label{TCAD Simulations}
The simulations presented in this paper were carried out using the Synopsys Sentaurus\footnote{http://www.synopsys.com} finite-element Technology Computer-Aided Design (TCAD) software framework.
\subsection{GaAs diode parameters and $CV$/$IV$ results}
\label{Diode simulation set-up}
\begin{figure}[tbp] 
\centering
\includegraphics[width=.47\textwidth]{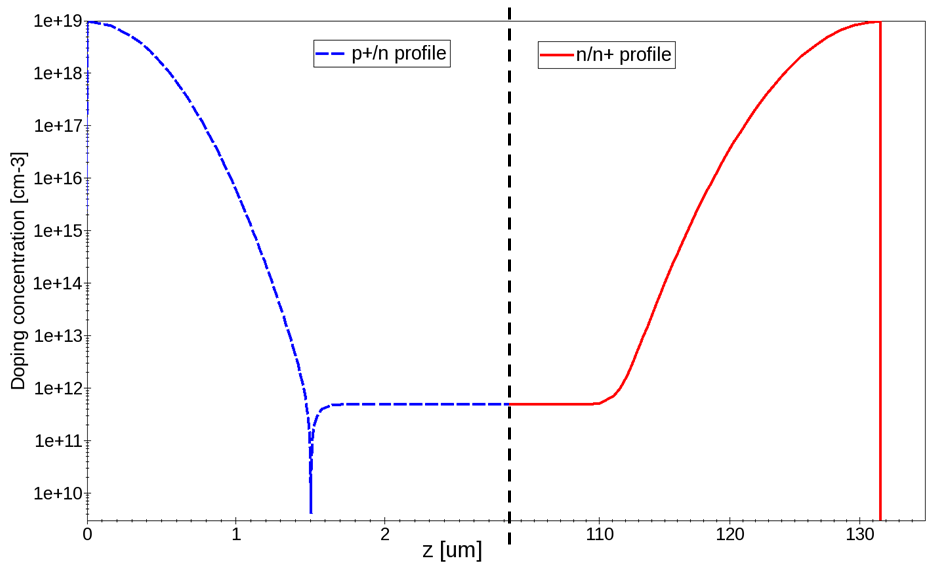}
\caption{Simulated doping profiles of p$^+$ and n$^+$ layers with tuned diffusion depths to create an effective structure similar to the real GaAs diode described in section~\ref{Processing of GaAs detectors}.} 
\label{fig:9}
\end{figure}

The initial simulated GaAs diode structure had the dimensions $A\times10\times131.5$ $\mu\textrm{m}^3$, where $A$ is the area factor to match the dimensions with the diode in figure~\ref{fig:5}. The thicknesses of the DC-coupled nickel contacts on the front and backplanes were 100 nm. Presented in figure~\ref{fig:9}, the heavily doped p$^+$ and n$^+$ implantations on the front and backplane, respectively, had the peak concentrations $1\times10^{19}$ $\textrm{cm}^{-3}$. The doping profiles were tuned to create approximations of the layer thicknesses presented in section~\ref{Processing of GaAs detectors}. 
The diode was biased by setting the electrode on the front surface to zero potential while the reverse voltage was provided from the backplane contact.

\begin{figure}[tbp] 
\centering
\includegraphics[width=.47\textwidth]{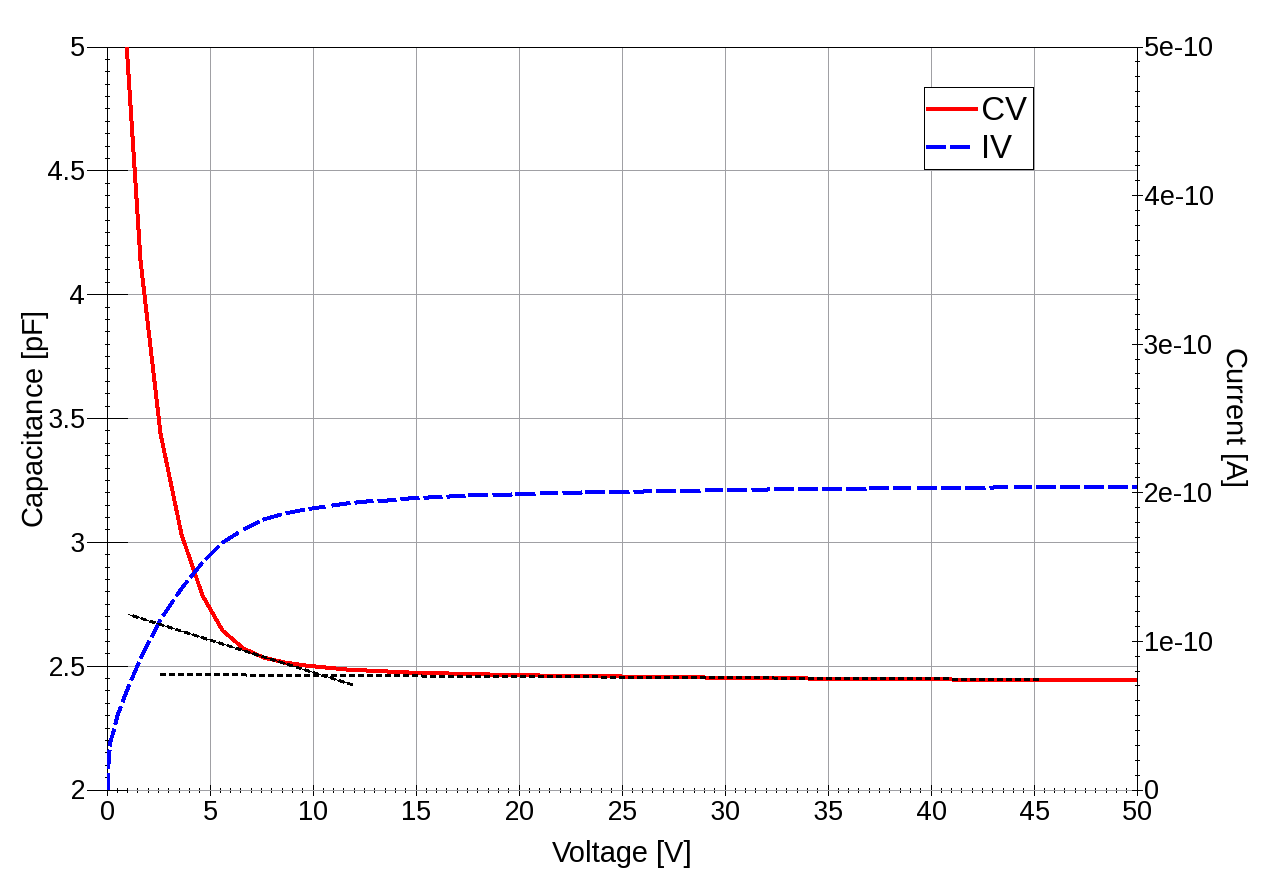}
\caption{Simulated $CV$ ($f$ = 10 kHz) and $IV$ curves at $T$ = 300 K for the measured GaAs diode parameters in section~\ref{Processing of GaAs detectors} with an active area thickness of 110 $\mu\textrm{m}$ and with the deep donor level from section~\ref{Deep level transient spectroscopy results}. 
The geometrical capacitance is $\sim$2.32 pF while the leakage current density is $\sim$8.5 nA/cm$^2$. The full depletion voltage of about 10 V was determined as in figure~\ref{fig:6a}.}
\label{fig:10}
\end{figure}

To reproduce the measured geometric capacitances, full depletion voltages and leakage current densities in figures~\ref{fig:6a} and \ref{fig:6b}, the deep donor level suggested in section~\ref{Deep level transient spectroscopy results} was implemented to the GaAs epi-layer. As can be seen from figure~\ref{fig:10} the simulated results are in line with the measurements after the capture cross section and concentration were tuned to $\sigma=5\times10^{-10}$ cm$^{2}$ and $N_{\textrm{t}}=1.4\times10^{12}$ cm$^{-3}$, respectively.
%
%
%
\subsection{Transient current simulation results}
\label{Transient current simulation results}

In GaAs and materials of similar band structure a negative differential mobility is generated by high driving fields. This effect is caused by the transfer of electrons into an energetically higher side valley with a considerably larger effective mass, i.e.  
when the electric field in the material reaches a threshold level, the mobility of electrons starts to decrease with increased electric field due to transferred electrons from one valley in the conduction band to another. In the process the electron effective mass moves from small to large and the mobility decreases from very high to very low \cite{Ruch1968,Dargys1994}. For the simulation of the effect the charge injection depth was set to $\sim$3 $\mu\textrm{m}$ from the device front surface to generate the transient signal essentially from the electron drift.

\begin{figure}[tbp] 
\centering
\includegraphics[width=.47\textwidth]{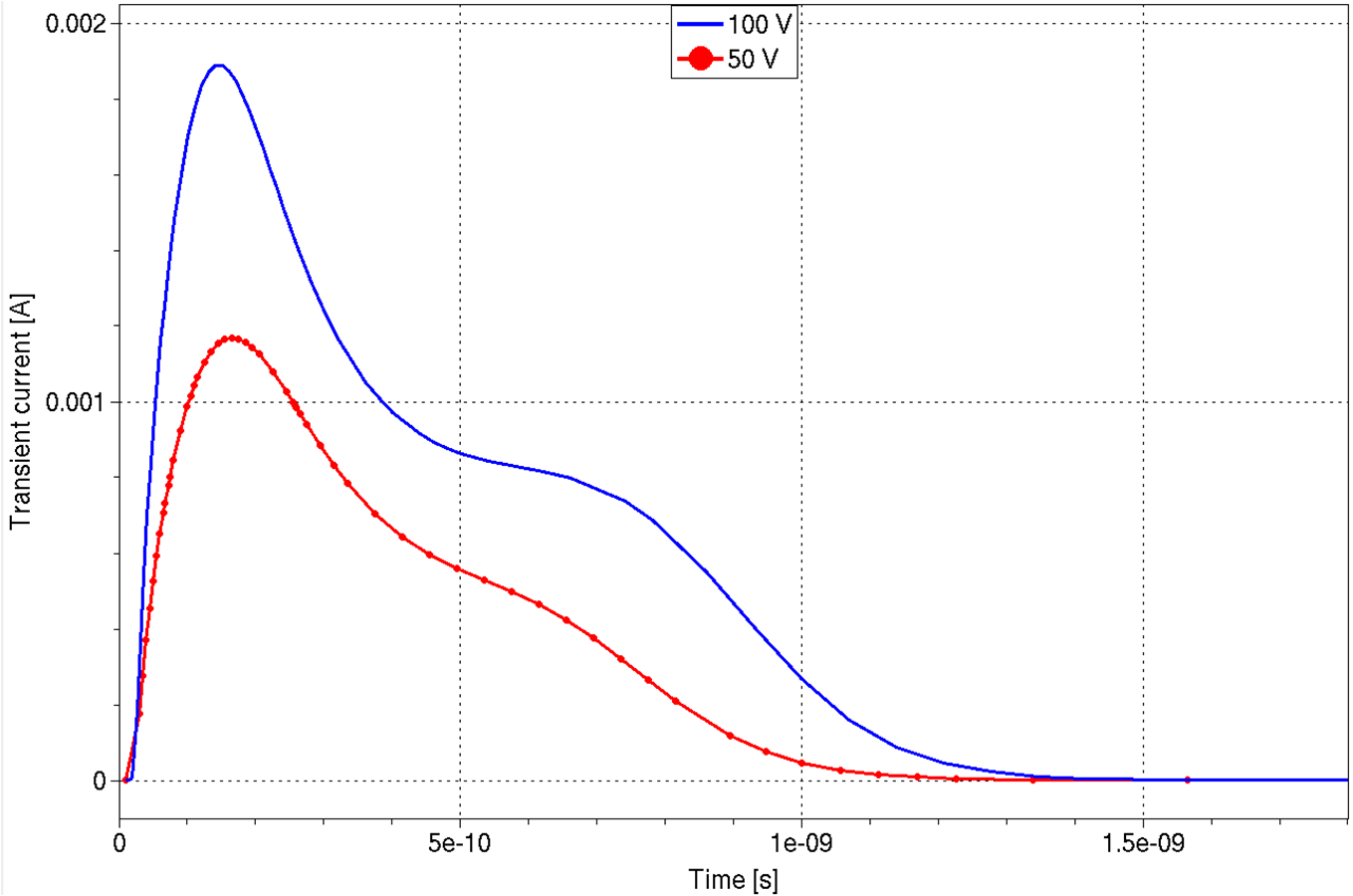}
\caption{Simulated transferred electron effect on transient current in GaAs diode. The shortest collection time is given by the voltage producing a drift velocity closest to the saturation value.}  
\label{fig:11}
\end{figure}

Illustrated in figure~\ref{fig:11} are the simulated transient currents 
in the GaAs diode at $T$ = 300 K. As can be seen, the macroscopic result of the transferred electron effect is reproduced, i.e. when the voltage and thus, the electric field in the diode moves closer to the values that produce the saturation of the electron drift velocity a reduced collection time $t_{\textrm{coll}}$ is observed. 
In the GaAs diode with an active volume thickness of about 110 $\mu\textrm{m}$ the voltage required for the saturation is $\sim$37 V. The simulated transient signal shapes and the difference in signal heights are in line with the measurements in figure~\ref{fig:7}. To reproduce the signal height differences agreeing with the measurement the simple diode structure described in section~\ref{Diode simulation set-up} had to be replaced by a larger structure that took into account the lateral expansion of the electric field beyond the diode region. If the lateral expansion of the electric field is not considered the fields for the given voltage will be higher than in the real diode, affecting the drift velocities. For simplicity, the deep donor level was not applied for the transient simulations that were carried out to investigate the transferred electron effect. Hence, the absolute collection times are shorter than the measured.  
%

\section{Conclusions}
\label{Conclusions}
X-ray detectors made on thick epitaxial GaAs were successfully processed at Micronova Nanofabrication Centre. The epitaxial GaAs layers were grown in a horizontal Chloride Vapor Phase Epitaxy (CVPE) reactor. A growth rate of undoped GaAs of about 10 $\mu\textrm{m}$/h was achieved and pin-diode devices having undoped layer thicknesses of 110 $\mu\textrm{m}$ and 130 $\mu\textrm{m}$ were grown. The finalized detectors were subjected to $CV$/$IV$, TCT and DLTS characterization. The full depletion voltage ($V_{\textrm{\small fd}}$) is in the range of 8-15 V and the leakage current density in the order of 10 nA/cm$^2$. According to the TCT measurements, the signal is collected faster than in 10 ns, while the DLTS analysis revealed a significant concentration of deep level electron traps in the epitaxial layer. 
TCAD simulations with an effective deep donor defect level reproduced the measured leakage current density and full depletion voltage. 
The transient simulations 
displayed the transferred electron effect producing similar behaviour with the measured TCT signals.

Thus, it can be concluded that these GaAs sensors are suitable for soft X-ray detection. There is ongoing work with a new vertical CVPE reactor to optimize the epitaxial process in terms of dislocation density as well as to better control the intrinsic stress causing wafer bow.


\bibliography{mybibfile}

\begin{thebibliography}{10}
\expandafter\ifx\csname url\endcsname\relax
  \def\url#1{\texttt{#1}}\fi
\expandafter\ifx\csname urlprefix\endcsname\relax\def\urlprefix{URL }\fi
\expandafter\ifx\csname href\endcsname\relax
  \def\href#1#2{#2} \def\path#1{#1}\fi

\bibitem{Owens2004}
A.~Owens, A.~Peacock, {C}ompound semiconductor radiation detectors, Nucl.
  Instr. $\&$ Meth. A 531~(1-2) (2004) 18--37.
\newblock \href {http://dx.doi.org/10.1016/j.nima.2004.05.071}
  {\path{doi:10.1016/j.nima.2004.05.071}}.

\bibitem{Sellin2006}
P.~Sellin, J.Vaitkus, {N}ew materials for radiation hard semiconductor
  detectors, Nucl. Instr. $\&$ Meth. A 557~(2) (2006) 479--489.
\newblock \href {http://dx.doi.org/10.1016/j.nima.2005.10.128}
  {\path{doi:10.1016/j.nima.2005.10.128}}.

\bibitem{Bhatta1997}
P.~Bhattacharya, Semiconductor Optoelectronic Devices, 2nd Edition, Prentice
  Hall Inc., 1997.

\bibitem{Ruch1968}
J.~G. Ruch, G.~S. Kino,
  \href{{http://journals.aps.org/pr/pdf/10.1103/PhysRev.174.921}}{{T}ransport
  properties of {GaA}s}, Phys.Rev. 174~(3) (1968) 921--931.
\newline\urlprefix\url{{http://journals.aps.org/pr/pdf/10.1103/PhysRev.174.921%
}}

\bibitem{Dargys1994}
A.~Dargys, J.~Kundrotas, Handbook on the physical properties of Ge, Si, GaAs
  and InP, 1st Edition, Science and Encyclopedia Publ., Vilnius, Lithuania,
  1994.

\bibitem{Sze1981}
S.~M. Sze, Physics of Semiconductor Devices, 2nd Edition, John Wiley \& Sons,
  New Jersey, 1981.

\bibitem{Kostamo2008}
P.~Kostamo, S.~Nenonen, S.~V{\"{a}}h{\"{a}}nen, L.~Tlustos, C.~Fr{\"{o}}jdh,
  M.~Campbell, Y.~Zhilyaev, H.~Lipsanen, {GaAs M}edipix2 hybrid pixel detector,
  Nucl. Instr. $\&$ Meth. A 591 (2008) 174--177.
\newblock \href {http://dx.doi.org/10.1016/j.nima.2008.03.050}
  {\path{doi:10.1016/j.nima.2008.03.050}}.

\bibitem{Tlustos2008}
L.~Tlustos, M.~Campbell, C.~Fr{\"{o}}jdh, P.~Kostamo, S.~Nenonen,
  {C}haracterisation of an epitaxial {GaAs/M}edipix2 detector using
  fluorescence photons, Nucl. Instr. $\&$ Meth. A 591 (2008) 42--45.
\newblock \href {http://dx.doi.org/10.1016/j.nima.2008.03.020}
  {\path{doi:10.1016/j.nima.2008.03.020}}.

\bibitem{Wu2014}
X.~Wu, et~al., {R}adiation detectors fabricated on high-purity {GaA}s epitaxial
  materials, JINST 9~(C12024).
\newblock \href {http://dx.doi.org/10.1088/1748-0221/9/12/C12024}
  {\path{doi:10.1088/1748-0221/9/12/C12024}}.

\bibitem{Eremin1996TCT}
V.~Eremin, N.~Strokan, E.~Verbitskaya, Z.~Li, {D}evelopment of transient
  current and charge techniques for the measurement of effective net
  concentration of ionized charges (${N}_{\textrm{eff}}$) in the space charge
  region of p-n junction detectors, Nucl. Instr. $\&$ Meth. A 372 (1996)
  388--398.
\newblock \href {http://dx.doi.org/10.1016/0168-9002(95)01295-8}
  {\path{doi:10.1016/0168-9002(95)01295-8}}.

\bibitem{Dingle1969}
R.~Dingle, {R}adiative {L}ifetimes of {D}onor-{A}cceptor {P}airs in p-{T}ype
  {G}allium {A}rsenide, Phys.Rev. 184~(3) (1969) 788--796.
\newblock \href {http://dx.doi.org/10.1103/PhysRev.184.788}
  {\path{doi:10.1103/PhysRev.184.788}}.

\end{thebibliography}

\end{document}